\documentclass[twocolumn]{aastex631}

\usepackage{mathtools, graphicx, booktabs, soul, multirow}

\usepackage{hyperref}
\hypersetup{
    colorlinks=true,
    linkcolor=blue,
    citecolor=blue,
    urlcolor=cyan,
}

\begin{document}

\title{The Cohesive Object Sequence: The Mass-Density Distribution of Astronomical Objects from Asteroids to Stars
\footnote{2026 Apr 6}}

\author[0000-0002-8482-4669]{Gabriel M Steward}
\affiliation{University of Idaho, Moscow, Idaho 83844}
\affiliation{Corresponding Author {\tt gsteward@uidaho.edu}}

\author[0000-0002-8592-0812]{Matthew Hedman}
\affiliation{University of Idaho, Moscow, Idaho 83844}
\accepted{by Publications of the Astronomical Society of the Pacific (PASP)}

\begin{abstract}

Plotting the mass-density of a wide range of astronomical objects as a function of their mass reveals that the vast majority of these objects fall along a ``cohesive object sequence'' that extends all the way from asteroids to the largest stars. Trends and features within this sequence reflect fundamental astronomical processes and phenomena, including the gravitational contraction of progressively higher-mass planets and the onset of nuclear reactions within stars. Meanwhile, compact stellar remnants fall well off this sequence, reflecting their extreme natures. This type of plot is therefore useful both for showcasing the relationships and connections between a wide range of astronomical objects and for clarifying the distinctions used to identify particular types of objects.

\end{abstract}


\section{Introduction} \label{sec:intro}

Trends among the measurable characteristics of astronomical objects have long been used to characterize and understand their properties. Most famously, the color-magnitude (or Hertzsprung-Russell) diagram displays the relationship between the color temperature and luminosity of stars and is used to define the main sequence and illustrate the life cycle of stars \citep{Carroll2017}. More recently, plots of the mass and radius for exoplanets have revealed distinct domains corresponding to different planetary types \citep{Chen2016, Muller2024}. Meanwhile, the distribution of mass densities for small bodies like asteroids \citep{Carry2012} provides information about the compositional diversity of the Solar System.

While asteroids, planets, and stars are often considered separately, there are objects like Pluto, Ceres, and brown dwarfs that blur the boundaries between them. Hence it is reasonable to treat them as parts of a larger group of astronomical objects that have well-defined masses and radii. Since there is not yet a generally accepted term for this broad class of objects, we will refer to them as ``cohesive objects'' here since they all have relatively well-defined surfaces due to direct interactions among their fundamental components. We deliberately exclude both nebulae and galaxies from this group, since these are systems with components (either atoms or stars) that are essentially collisionless and only interact via gravity (while characteristic sizes and masses of these sorts of systems can be defined, they all lack the relatively well-defined surfaces associated with objects like planets and stars). 

Our definition of cohesive object is extremely broad, including asteroids, comets, dwarf planets, moons, planets, brown dwarfs, stars, white dwarfs, neutron stars, and even black holes. We choose to include black holes as part of this class even though the event horizon is not a boundary of physical matter (although there are debates on this matter, see \cite{Mann2022}). This is because black holes are certainly singular objects with a well-defined physical size given by their Schwarzschild event horizons. 

A large number and wide variety of cohesive objects have had their mass and radius measured over the years. Furthermore, with just these two properties, we can also determine physically relevant quantities such as density and surface gravity. This means that we can generate plots of these parameters for each of these objects, providing a visual representation of trends among objects ranging from asteroids to giant stars.

In this paper, we present graphs of the mass-density relation for all cohesive objects with decent mass and radius measurements, ranging from minuscule asteroids to supermassive black holes. These plots provide a way to examine the overall distribution of objects in the universe, connecting many disciplines across astrophysics. Like the HR diagram, this distribution plot suggests many clear divisions by which to categorize objects, and even has a "main sequence" of sorts on which the vast majority of objects fall, which we call the ``cohesive object sequence.'' 

We begin by describing our data collection and selection process in Section \ref{sec:methods} and explore the data set in Section \ref{sec:results}, where we sketch out classification implications, outlier examinations, and connections between seemingly distinct objects. Section \ref{sec:conc}, our conclusion, follows from there.

\section{Methods} \label{sec:methods}

\begin{table*}[p] 
\centering 
\begin{tabular}{cc}
\multicolumn{2}{c}{\textbf{Table 1}} \\
\multicolumn{2}{c}{Cohesive Object Sources} \\
\hline
\hline
Cohesive Object Class & Sources \\
\hline
\hline
\multirow{2}{3.5em}{Asteroids} & \cite{Carry2012}, \cite{Lauretta2019}, \cite{Russell2012}, \\
& \cite{Vernazza2021}, and \cite{Watanabe2019} \\  
\hline
Black Holes & \cite{EHTC2019}, \cite{Tamburini2019}, \cite{EHTC2022}, and \cite{Daly2023} \\
\hline
Brown Dwarfs & \cite{Grieves2021}, \cite{Limbach2024}, and \cite{Stassun2007} \\
\hline
Comets & \cite{Carry2012} and \cite{Patzold2016} \\
\hline
Earth & \cite{Archinal2018} and \cite{Folkner2009} \\
\hline
Exoplanets & \cite{Kanodia2024} and \cite{NASAExoplanetArchive} \\
\hline
Jovian Moons & \cite{Anderson2005}, \cite{Archinal2018}, and \cite{Bagenal2007} \\
\hline
Jupiter & \cite{Archinal2018} and \cite{Bagenal2007} \\
\hline
Mars & \cite{Archinal2018} and \cite{Bills2005} \\
\hline
Martian Moons & \cite{Archinal2018} and \cite{Bills2005} \\
\hline
Mercury & \cite{Archinal2018} and \cite{Anderson1987} \\
\hline
The Moon & \cite{Archinal2018}, \cite{Goossens2016}, and \cite{Lemoine2014} \\
\hline
Neptune & \cite{Archinal2018} and \cite{Jacobson2009} \\
\hline
\multirow{2}{3.5em}{Neutron Stars} & \cite{Doroshenko2022}, \cite{GonzalezCaniulef2019}, \\
& \cite{Nattila2017}, \cite{Reardon2015}, and \cite{Riley2019} \\
\hline
Saturn & \cite{Archinal2018} and \cite{Jacobson2006} \\
\hline
Saturnian Moons & \cite{Archinal2018}, \cite{Jacobson2006}, \cite{Jacobson2022}, and \cite{Thomas2020} \\
\hline
\multirow{2}{2em}{Stars} & \cite{Boetticher2017}, \cite{Bond2017}, \cite{Grieves2021}, \\ & \cite{Morin2010}, \cite{Pineda2021}, \cite{Southworth2014}, and \cite{Torres2009} \\
\hline
The Sun & \cite{Emilio2012} and \cite{Park2021} \\
\hline
\multirow{3}{10.1em}{Trans-Neptunian Objects} & \cite{Brozovic2015}, \cite{Brown2017}, \cite{Carry2012}, \cite{Holler2021}, \\
& \cite{Kiss2019}, \cite{Nimmo2017}, \cite{Ortiz2012}, \cite{Ortiz2017}, \cite{Ragozzine2009}, \\
& \cite{Sicardy2011}, \cite{Souami2020}, and \cite{Stern2018} \\
\hline
Triton & \cite{Jacobson2009} and \cite{Thomas2000} \\
\hline
Uranian Moons & \cite{Jacobson2014} and \cite{Thomas1988} \\
\hline
Uranus & \cite{Archinal2018} and \cite{Jacobson2014} \\
\hline
Venus & \cite{Archinal2018} and \cite{Konopliv1999} \\
\hline
White Dwarfs & \cite{Parsons2017} and \cite{Bond2017} \\
\hline
\end{tabular}
\caption{Sources used for the various classes of cohesive object. The final data set includes mass, radius, and errors for both; these values were sometimes calculated from values within the sources rather than provided directly.}
\label{table:1}
\end{table*}

Masses and radii for various types of cohesive objects were gathered from the sources tabulated in Table \ref{table:1}. We only included objects that had mass and radius measurements with reported errors, or objects where mass and radius could be readily calculated from other reported values. We included objects where the measurements had asymmetric error bars and where the mass and radius data came from different sources. 

Errors on derived quantities like density have error bars propagated in the standard way. For irregular objects like asteroids, comets, and small moons, we use the ``effective radius", which is the radius of a sphere with the same volume as the object. Note that for this particular survey, we focus on a subset of measurements with relative mass errors better than 0.5 and relative volume errors better than 0.5 (that is, relative radius errors better than about 1/6). The value of 0.5 is chosen arbitrarily, but the choice of assigning it to mass and volume rather than mass and radius is so that the density error is equally weighted between the mass and volume.

The parameters for exoplanets were taken from the NASA Exoplanet Archive (called on 2025 April 15) \citep{NASAExoplanetArchive}. We first removed from this database all entries that lacked recorded mass, mass errors, radii, and radii errors. From these, we only included the most recently updated entries for each individual planet. If several entries had the same update date, we chose the one with the largest relative error in the radius or mass. Although the choice was somewhat arbitrary, the less certain value likely made the fewest assumptions about the object. Only after this initial sorting did we remove entries with relative errors greater than 0.5.

Unfortunately, there are not similarly comprehensive databases for Solar System objects or stars, so we compiled those data from multiple sources, tabulated in Table \ref{table:1}. A few of the larger sources for stars include \cite{Southworth2014}, \cite{Torres2009}, and \cite{Pineda2021}. \cite{Carry2012} provided many of the asteroids and small objects.

We were only able to gather four measurements of neutron star radii sufficient for our purposes. In the specific case of PSR J0030+0451, two different groups have papers with different measurements and identical release dates \citep{Miller2019, Riley2019}. We selected \cite{Riley2019} for our data set following the procedure outlined above for selecting NASA Exoplanet Archive data.

Only one black hole's radius has been satisfactorily measured thus far: M87*, the subject of the famous Event Horizon Telescope image \citep{EHTC2019}. The black hole in the center of our own galaxy, Sagittarius A*, has also been measured \citep{EHTC2022}, but the volume errors exceed 0.5; in the interests of giving M87* some context, we have retained Sagittarius A* in our final data set despite this. 

Notably, determining the effective radii of black hole event horizons is complicated by the fact that they deform considerably when they have high angular momentum, so a true deduction of the event horizon radius requires knowing how much they spin. We obtain these values from \cite{Tamburini2019} and \cite{Daly2023}.  We obtain our final radius values assuming a Kerr metric black hole \citep{Visser2007}.

Mass for Solar System objects is often reported in terms of GM with high precision for mission and observation planning. However, we sought actual mass, and so we often calculated new errors from the uncertainty in $G$ itself \citep{Gillies1997}. Occasionally, a direct mass would be reported that had relative errors smaller than the uncertainty in $G$. In such cases, we truncated the relative errors to those of $G$, as anything more precise was not believable. 

We checked for obvious outliers for each object type, and removed measurements that had clear issues (e.g. a mass that was for a binary rather than a single object). However, outlier objects with no obvious issues, like high-density exoplanets and asteroids, were retained.

In the end, 2157 objects yielded suitable mass and radius estimates for this study. This dataset is available at the GitHub repository \url{https://github.com/GMBlackjack/CohesiveObjects/tree/main}. 

\section{Results and Discussion} \label{sec:results}

\begin{figure*}[htbp]
\centering
\includegraphics[scale = 0.35]{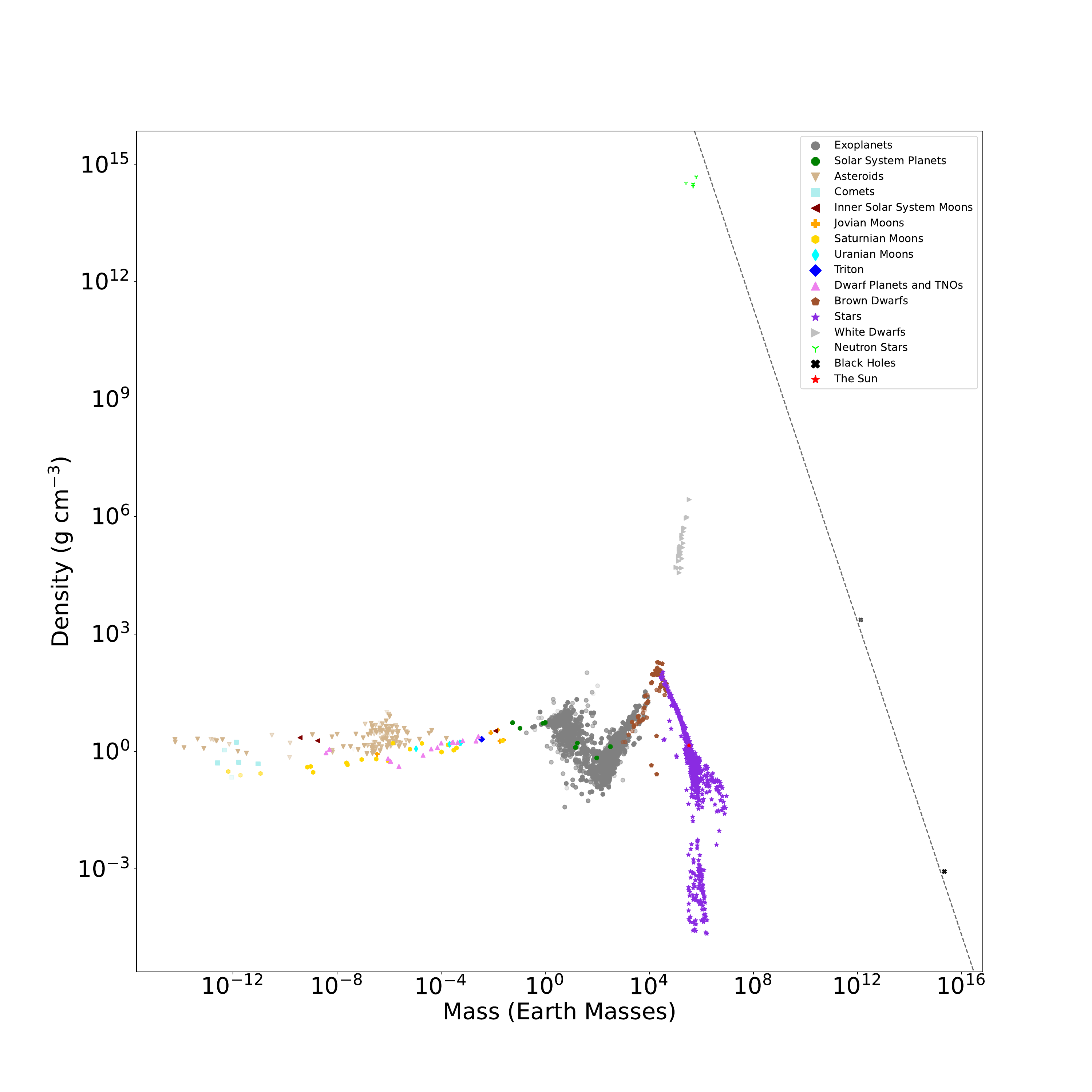}
\centering
\caption{The relationship between density (in g cm$^{-3}$) and mass (in Earth masses) for cohesive objects in the universe. Symbol colors and shapes indicate the kind of object. The transparency of each point represents how large the errors are: the more transparent objects have larger errors with a maximum relative permitted error of 0.5 in either mass or volume, whichever was larger for the object in question. No object has no error, but anything less than 0.01 is largely indistinguishable from a solid point. The dashed line represents the Schwarzschild black hole event horizon.}
\label{fig:1}
\end{figure*}

Figure \ref{fig:1} shows the density of all the different types of cohesive objects we considered as a function of their mass, which we think provides the clearest information about trends in the properties of these objects.

\begin{figure*}[htbp]
\centering
\includegraphics[scale = 0.45]{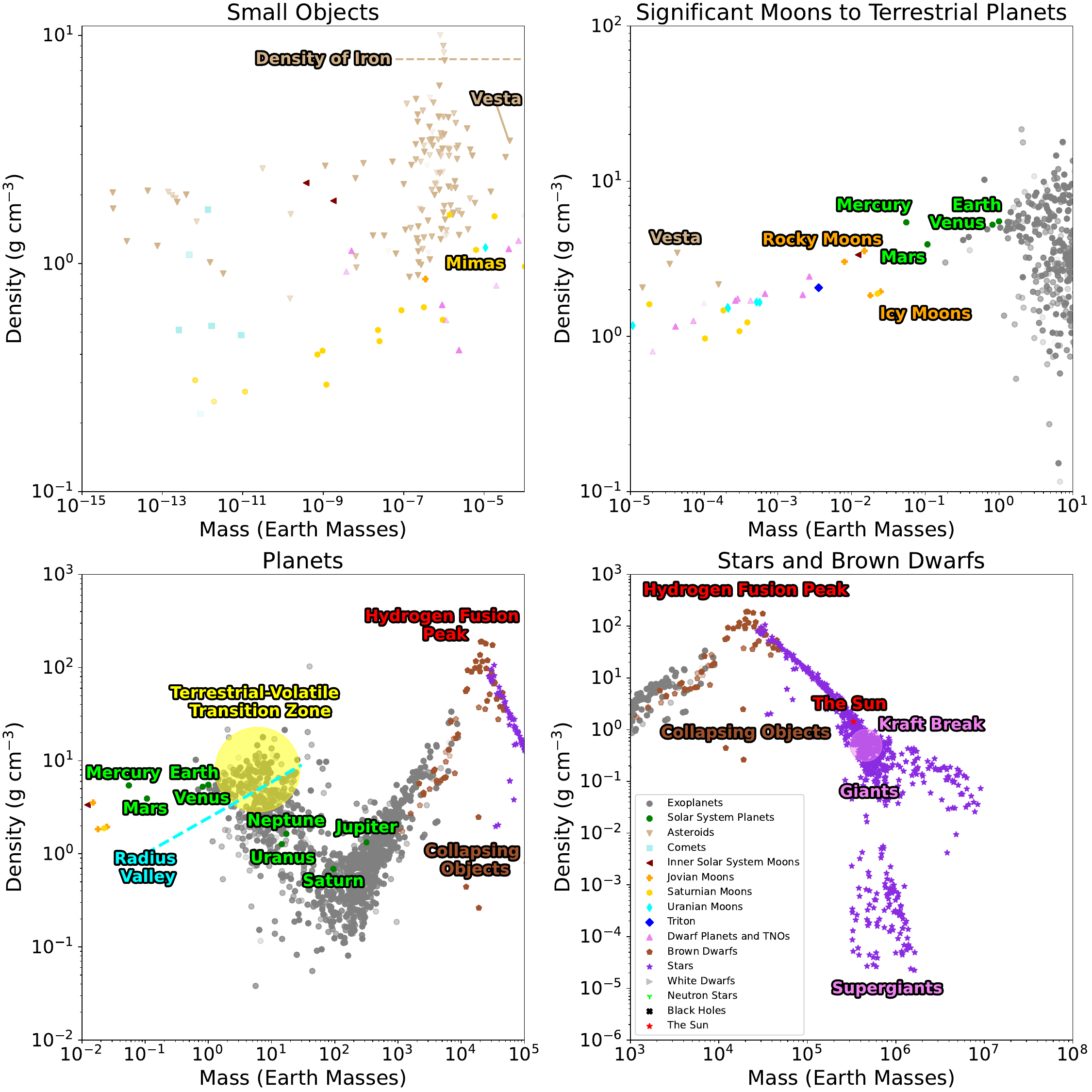}
\centering
\caption{Zoomed in views of the cohesive object sequence shown in Figure \ref{fig:1}. Objects and regions of note discussed in the main text are labeled. The boundaries of the labeled regions are not intended to be definite, and are merely generalizations. The few asteroids above the density of iron are thought to be physically impossible, likely due to erroneous measurements. Mimas is the smallest gravitationally rounded object, and Vesta is the largest irregular object; the two of them mark the transition between irregular and rounded objects. In the small-mass regime, there are no special labels for ``Asteroids" or ``Comets," as their populations blend together alongside various small moons. The major rocky and icy moons are highlighted to show that a clear distinction in composition is present at that point in the plot, though shortly thereafter, such distinctions vanish. The Solar System's planets and the sun are labeled primarily for context. Various transition points along the cohesive object sequence are marked, including the sharp hydrogen fusion peak where stars are created, the less clear but still identifiable kraft break where stars change from outwardly radiative to convective, and the very poorly constrained terrestrial-volatile transition zone where, somewhat annoyingly, most of the planets in the universe seem to reside. The radius valley, a location of very few observed exoplanets, cuts right across this nebulous region. The different regimes of exoplanets (terrestrial, volatile, gaseous) and the main sequence stars are not labeled as they are ``standard" parts of the cohesive object sequence; this is opposed to giant and supergiant stars, which divert from it. Between gas giants and stars, there is a small population of collapsing objects, sparse simply due to the fact that objects do not stay in this state very long, quickly moving up the plot with density to enter the cohesive object sequence.} 
\label{fig:2}
\end{figure*}

\begin{figure*}[htbp]
\centering
\includegraphics[scale = 1]{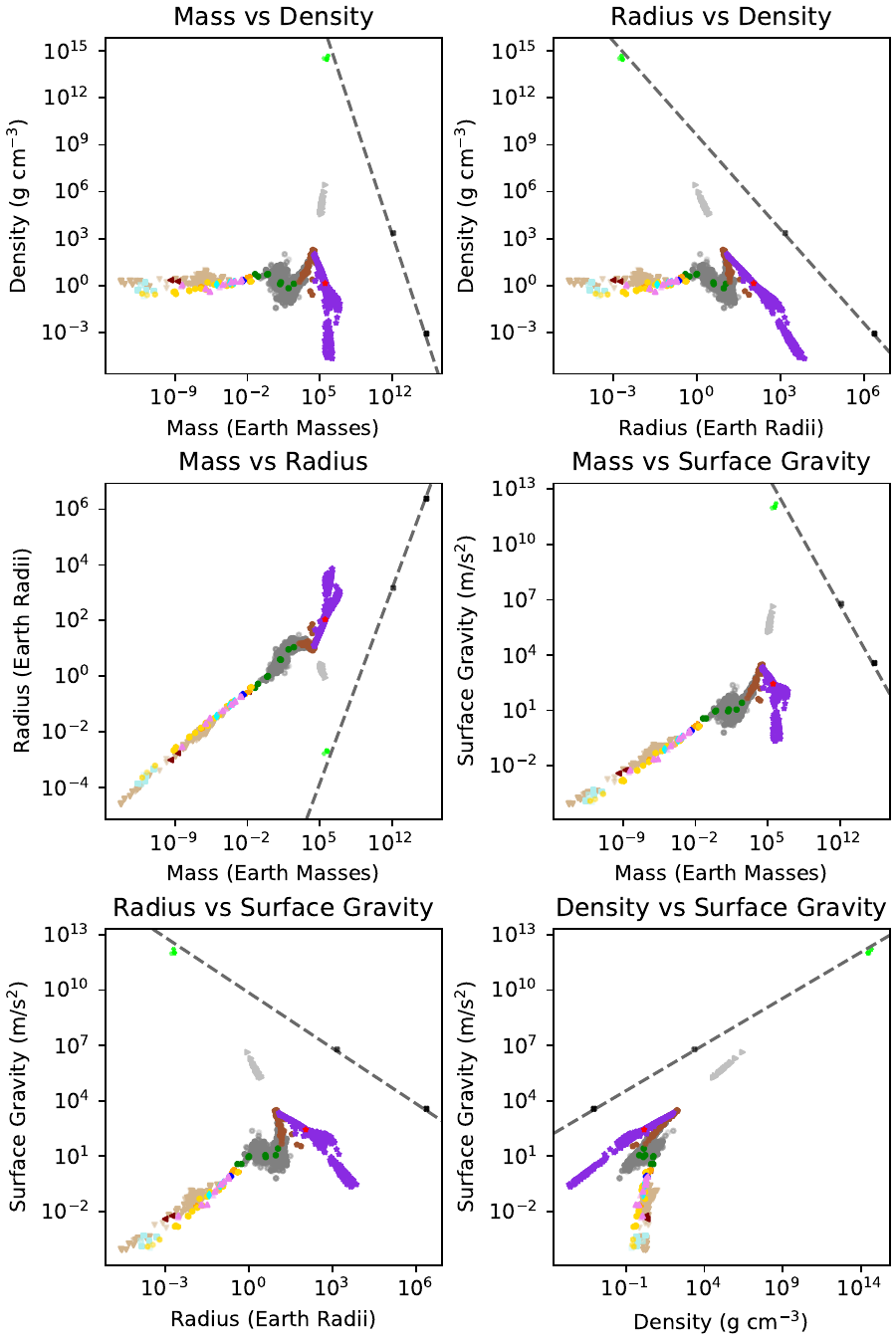}
\centering
\caption{Alternate views of the data set showing different combinations of mass, radius, density, and surface gravity. Colors and shapes are identical to Figure \ref{fig:1}.}
\label{fig:3}
\end{figure*}

Expanded views of various sections of this plot are shown in Figure \ref{fig:2}, allowing individual objects to be identified among the otherwise dense clouds of points. For the sake of completeness, we also provide plots of radius, mass, density, and surface gravity in Figure \ref{fig:3}.

The most remarkable aspect of Figure~\ref{fig:1} is that it showcases an unbroken progression from the smallest asteroids to the largest stars, spanning most other kinds of astronomical objects over 12 orders of magnitude in mass when disregarding the black holes, as that region is undersampled. We call this progression the ``cohesive object sequence,'' after the stellar main sequence, which it contains. Of the objects considered here, only compact stellar remnants, giant stars, and certain actively collapsing bodies clearly lie off the cohesive object sequence. 

One might be tempted to describe the cohesive object sequence as a representation of how objects develop as they accrete more mass, but such an interpretation would be inconsistent with how these objects actually form. While terrestrial planets do form by accumulating smaller rocky objects, stars form from clouds of gas and have formed since before there were enough heavy elements to even make asteroids \citep{Maio2010}. Formation pathways for objects of the same class can vary as well: larger gas giants could form in a circumstellar disc or directly from a stellar nursery \citep{Schlaufman2018}. Instead, this sequence's features say more about how the composition and structure of these objects varies with mass. For example, there are no rocky bodies the size of Jupiter, and there are no gaseous objects the size of asteroids. This is similar to the stellar main sequence, which says little about stellar formation processes but rather gives insights into their equilibrium properties.    

Another key insight is that there are no huge gaps in the sequence, with differences in the density of points primarily due to the limitations of the relevant observations. Cohesive objects, therefore, cannot be easily divided into distinct groups based solely on their mass or radius. However, there are multiple kinks and bends in the sequence that provide a physically motivated way to categorize objects. These kinks persist even when different parameters are plotted against each other, as seen in Figure \ref{fig:3}. 

The lack of clear gaps in this context means that there are guaranteed to be objects in transition regions that would be difficult to classify as being one category or another--many objects sit right in the divide between terrestrial planets and volatile worlds (also known as ice giants), for instance.

Turning to a more detailed examination of the smaller-scale features in the cohesive object sequence, we begin by considering low-mass objects and work towards progressively larger objects. Of course, the relatively small number of objects with masses below $10^{-8}$ Earth masses is an observation bias due to larger objects being easier to observe and our decision not to include "tiny" objects such as meteoroids and interplanetary dust. Most of the objects at the extreme left side of the plot were visited by spacecraft, including the smallest object in our dataset, Itokawa \citep{Carry2012}.

The majority of small ($< 0.1$ Earth mass) objects cluster around $10^{-6}$ Earth masses, and here we can clearly identify objects with a wide range of densities due to them having different compositions and porosities. Most of the asteroids have generally larger densities than the trans-Neptunian objects and the moons in the outer Solar System, which is consistent with the asteroids being composed primarily of rock while the Trans-Neptunian Objects and outer Solar System satellites are composed primarily of ice. Many of the icy objects have densities below the expected densities of solid rock (3 g cm$^{-3}$) or ice (1 g cm$^{-3}$), likely because they are loose aggregates with significant porosity. Icy objects with minimal porosity and rocky objects with high porosity exist, as well as objects of mixed composition, which is why the rocky and icy object distributions bleed into each other rather than having any clear boundaries.

The densest asteroids have densities slightly higher than pure iron, and so likely represent inaccurate measurements. However, there are no doubt some true metallic meteorites with densities close to these values, due to fragmentation of differentiated bodies \citep{Scott2020}. 

After an object obtains enough mass, it will begin to gravitationally round itself \citep{Stern2002, Lineweaver2010}. The lowest-mass object that is known to be gravitationally rounded in a normal manner is Saturn's icy moon Mimas at $6.3\times10^{-6}$ Earth Masses \citep{Jacobson2006}, while the most massive irregular object is the asteroid Vesta at $4.3\times10^{-5}$ Earth masses \citep{Russell2012}. These two objects only differ in mass by a single order of magnitude, and so this region marks the relatively narrow transition between irregular and spherical objects. This trend coincides with the range of densities becoming increasingly narrow with increasing mass between 10$^{-6}$ and 10$^{-4}$ Earth Masses. There are no very dense asteroids of this size, but also no icy objects with particularly low densities. The trend manifests in a vaguely cone-like shape. The trend of increasing density in ice-rich Trans-Neptunian Objects with increasing mass has been noticed before by \cite{Bierson2019}, who attributed it to gravitational compression and a subsequent loss of porosity, a process which no doubt applies to rubble-pile asteroids as well. On the other hand, the lack of higher-density objects at higher masses requires another explanation. It may be that larger objects are more likely to have average material compositions, as it becomes more difficult to selectively accrete certain materials, or it could be that no object large enough to fragment into high-mass high-density fragments ever formed in the Solar System. This suggests that the cohesive object sequence may look different at this scale in other Solar Systems that experienced more or less violent histories. 

Between 10$^{-4}$ and 0.1 Earth masses, we reach the part of the sequence with the lowest number of known objects, which include the largest dwarf planets, major moons of the Solar System, and the smallest planets. The zone below 10$^{-3}$ Earth masses consists almost exclusively of ice-rich worlds. Pluto, Eris, and Triton cluster together between 10$^{-3}$ and 10$^{-2}$ Earth Masses, which is in line with Triton's suspected origin as a captured dwarf planet \citep{Agnor2006}.

Around 10$^{-2}$ Earth masses, we can see the data split into two trends. Io, Europa, and the Moon all fall in a line because they are largely rocky bodies, while Titan, Ganymede, and Callisto are in a separate line at lower densities due to being icy bodies. Mercury and Mars sit alone with few neighbors, since we are only now reaching the sensitivity required to detect these objects in other star systems. (Most small exoplanet candidates did not pass through the error requirements we placed on our data.)

However, it is still worth noting that Mars' density lies along a trend consistent with Venus, Earth, and other comparably-sized terrestrial exoplanets, while Mercury's density is well above this trend.  This is consistent with Mercury having an unusually high metal content \citep{Spohn2001}. It remains to be seen how often such particularly dense small planets occur in the universe.

This region of the plot also illustrates why it has been so difficult to define a lower mass threshold for a planet \citep{Stern2002, Soter2006, Metzger2022}. There are no obvious gaps in the object distribution below 1 Earth mass, and there are only relatively subtle and gradual transitions in density with mass throughout this range. The transition from asteroids to Earth-like objects therefore appears to be smooth with no clear features that would physically motivate a category boundary.

By contrast, there are much more dramatic features in this sequence above 1 Earth mass. First of all, objects have a far larger variety of densities above a few Earth Masses than they do between 0.1 and 1 Earth Masses. This marks the boundary between terrestrial planets and volatile-rich worlds like Uranus and Neptune. 

It is unclear where terrestrial planets end and volatile-rich worlds begin. There may also be a handful of "water worlds" that exist in this transition region \citep{Adams2008, Zeng2019}, a type of planet that is neither terrestrial nor volatile--at least not in the same sense that Uranus and Neptune are volatile. The unfortunate reality is that planets in this uncertain range are extremely common, and we do not have any examples of a planet in this regime within the Solar System. 

The true nature of this transition zone is an area of active research \citep{Muller2024, Lange2024, Zeng2019, Armstrong2019, Eylen2018}. However, it is notable that this region appears to be unique in that it contains a large range of cohesive object compositions over a relatively small range of masses. In general, it appears that smaller objects (around 3 Earth Masses) in this region tend to be more volatile-poor and higher density, while larger objects (around 30 Earth Masses) are more volatile-rich and lower density. However, there are also a handful of "overdense" worlds throughout this region that have densities above 3 g cm$^{-3}$ that may imply a rocky (or even highly metallic) composition, but there is currently no consensus as to what exactly these overdense worlds are \citep{Armstrong2020, Naponiello2023, Lange2024}. 

The range and variations in the densities across this region probably reflect limits on the conditions under which planets can typically form, though specific speculations are beyond the scope of this paper. It is also worth noting that this transition region is largely continuous and does not include any extensive break in the sequence. Even the so-called ``radius valley'' \citep{Eylen2018, Armstrong2019}, which can be seen in Figure \ref{fig:2}'s zoom on the planet region, is not a true divide, as there are some objects that straddle it. 

Leaving the ambiguous transition zone, we enter the realm of the volatile-rich worlds (objects also known as ice giants). Previously, as worlds gained more mass, their density increased; but here the trend is reversed, with higher masses producing lower densities. Uranus and Neptune sit in the middle of this region, completely average among others of their kind. It is worth noting that the volatile worlds themselves have the largest internal variation of any category on the entire cohesive object sequence. Almost everything else lies on narrow lines or at least regions with well-defined edges. Meanwhile, the volatile worlds have members of extremely high and low density across their entire mass range. 

There is a clear feature in the sequence around 100 Earth masses. At this location, not only does density start increasing with mass again, but also the variations in density begin to become narrower. This dramatic change in behavior marks the transition between volatile-rich worlds with a range of compositions and the gas giants composed primarily of hydrogen and helium, like Saturn and Jupiter.

It just so happens that Saturn sits keenly near the middle of this feature \citep{Helled2023}, indicating that it is one of the smallest gas giants possible, or perhaps even a member of a thin transitional zone. Below Saturn and Jupiter are a handful of volatile worlds and gas giants that are underdense--including planets inflated due to proximity to their stars \citep{Batygin2011}, and the so-called ``super-puffs'' \citep{Lee2016, Libby-Roberts2020}. By contrast, the cloud of overdense worlds cuts off near the masses of the smallest gas giants, suggesting that it must be extremely difficult to form an object so massive without accumulating a gaseous envelope. This makes sense, given that planets forming in a gas-rich disc are expected to experience runaway gas accretion when they grow large enough \citep{Bodenheimer1986, Venturini2016}. 

Between 100 and 10,000 Earth masses, the density steadily increases with increasing object mass. This reflects the fact that as hydrogen-rich objects like gas giant planets and brown dwarfs get more massive, they become denser with minimal change in physical size on a relatively predictable trajectory until they reach the masses required to start hydrogen fusion \citep{Baraffe1998, Boetticher2017}. In this context, it is interesting to consider the distinction between gas giants and brown dwarfs. No matter which parameters are plotted, there is no obvious feature in the sequence that can serve to clearly define the boundary between gas giants and brown dwarfs. Indeed, brown dwarfs and gas giants are only colored differently in Figure \ref{fig:1} because of the sources they were drawn from, and there is a large overlap between the two sets of objects. The current default differentiator between giant planets and brown dwarfs is the deuterium fusion burning limit \citep{Boss2005}, but we see no indication of a major change in these objects at that point. The actual effects of deuterium fusion on an object are relatively minor: it takes longer for objects undergoing the process to cool after formation \citep{Allen2003}, and their deuterium will be depleted, but in the end the object will stop burning and become inert over relatively small timescales of a few tens of millions of years \citep{Palla2005}. 

Our plot shows a paucity of mid-range brown dwarfs (at around $2 \times 10^4$ Earth masses). It is unclear if this is due to an observation bias, or if mid-range brown dwarfs are truly uncommon. The brown dwarfs that do exist in this region preclude it from being used as an easy delineator between gas giants and brown dwarfs, however.

Now that we've reached the definite end of all planets, it's worth taking another look at this category of objects. In every view of mass, radius, density, and surface gravity, there are clearly at least three separate regimes for planets: terrestrial planets, volatile-rich worlds, and gas giants. The clear distinctions between these three classes (which are about as dramatic as the distinction between planets and hydrogen-burning stars) call into question whether it is sensible to consider objects as different as Jupiter, Uranus, and Earth as the same basic type of object. Instead, perhaps it would be better to consider terrestrial planets, volatile-rich worlds, and gas giants as their own distinct groups. 

Our sequence ends with the stellar main sequence, which shows a clear trend of decreasing density with increasing mass. Low-mass (less than 10$^5$ Earth Masses) stars exhibit little variation in density, while higher-mass stars show larger variability. Examination of the stars section in Figure \ref{fig:2} shows a slight change in the mass-density slope between low-mass and high-mass stars at around 10$^5$ Earth masses. This has been observed before \citep{Gimnez1985, Camposeo2025} and corresponds with the "Kraft Break" where stars switch from outwardly convective to outwardly radiative \citep{Beyer2024}.

The Sun is a completely normal member of the main sequence. The giant branch breaks off from the stellar main sequence (and simultaneously the cohesive object sequence) at masses around that of the Sun and points toward the disconnected supergiant region. The stellar main sequence ends with blue stars around 10$^7$ Earth masses, and the cohesive object sequence does not carry it into any further objects. Perhaps population III stars would extend the sequence further, but these have not yet been conclusively observed \citep{Larkin2022}.

There are a handful of objects not on the cohesive object sequence. Naturally, there are the giant and supergiant stars approaching the ends of their lives as they run out of nuclear fuel to burn in their cores. With densities getting well below 0.01 g cm$^{-3}$, the supergiant stars include the least dense cohesive objects in our dataset.

Finally, we can note that all of the compact objects--white dwarfs, neutron stars, and black holes--are also not on the cohesive object sequence, and the jarring lack of objects connecting these regimes to the sequence is indicative of the dramatic processes that form them. While there is no "gradient" to becoming a compact object, it is worth noting that the distribution of white dwarfs does vaguely point away from the stars and toward neutron stars, and neutron stars fall close to the line that corresponds to black holes. The two black holes we have plotted lie slightly above the Schwarzschild radius line. Both are spinning at near the maximum allowed angular momentum \citep{Tamburini2019, Daly2023}, which corresponds to smaller event horizons \citep{Visser2007}, and thus higher densities.

The only other distinct class of object clearly off the cohesive object sequence are collapsing objects: very young stars, brown dwarfs, and gas giants that are still collapsing \citep{Hartmann2016, Muller2021}. We do not have many of these objects in our dataset because so few of them have been measured, much like there are not many known stars that are currently in the process of moving onto the supergiant stage of their lives. Objects simply do not spend enough time in either of these regimes for there to be a large population of them. 

\section{Conclusion} \label{sec:conc}

We have collected data on over two thousand cohesive objects--astronomical objects made of components in physical contact with each other. All of these objects share some basic parameters: mass, radius, density, and surface gravity, which are used to compare trends between them. We find that most cohesive objects in the universe follow a cohesive object sequence that shows continuous connections from asteroids all the way to the largest stars, with only a handful of extreme objects lying off this sequence. The final resulting graphs provide insights into trends among the universe's objects, highlighting connections between objects of very different sizes that are often discussed separately, as well as suggesting new ways to classify and distinguish between different object types. 

Some plots similar to ours have been made in the past, though most focus on a narrow section of the plot \citep{Chen2016, Muller2024} or try to combine everything ever considered in the universe together to the point of obscuring interrelating trends \citep{Lineweaver2023}. One exception is \citep{Enoto2021}, which has a variety of real object types plotted alongside theoretical distributions, ranging from the Earth to compact objects. Our work expands theirs considerably, particularly with objects smaller than Earth, and we fill out regimes where \cite{Enoto2021} only have a handful of real data points surrounded by a theoretical boundary region or evolution line. Our primary result in Figure \ref{fig:1} deliberately leaves off labels from theory, focusing exclusively on real objects to give an unbiased view of the universe. The large number of data points provides a sanity check for the cohesive object sequence: we generally don't have to "trace" lines with our eyes; we simply see the objects where they are.

Previous work similar to ours often plots the data as mass-radius (\cite{Chen2016} and \cite{Enoto2021}, though \cite{Muller2024} plots multiple variables). The mass-radius choice is the simplest one, as these are usually the two parameters most readily measured, but we find it much easier to see delineations in object type through mass-density plots that adjust the slope between positive and negative values, as opposed to the near-continual ascending slope of a mass-radius plot, as can be seen in Figure \ref{fig:3}.

Our ultimate goal is that our mass-radius plot, much like the HR diagram, will be used broadly to showcase the relations between astrophysical objects in a succinct, easy-to-understand manner. Our plots show real distributions of objects, and these can be compared with theoretical distributions as a quick check on accuracy--and, if a model predicts a type of object not represented in the data set, it will lead us to ask if that is due to a model deficiency or a lack of observations in that data regime. We wonder if a model of the entire cohesive object sequence could be made that captures the blend from asteroids all the way to the heaviest stars. 

Furthermore, we wish to encourage collaboration between different research fields that usually study specific objects in isolation. Most striking of these connections is, we believe, the unbroken path from asteroids to terrestrial planets. There is no sharp distinction or change of character from the smallest objects to objects like Earth on the cohesive object sequence, save for the limiting of density variation. In addition, many objects on vastly different sections of the cohesive object sequence have similar densities and surface gravities--a clue that, perhaps, similar mathematical models could apply to multiple types of objects. Is there any reasonable connection between the densities of the smallest stars and the overdense exoplanets? Can anything be drawn from the similar distributions of surface gravity between volatile worlds and main-sequence stars? Such environments are usually considered radically different, and yet, through our work, we see reason to compare them. We encourage readers to come up with more questions for rarely compared environments that we haven't even considered.

We hope that plots like this can help with efforts to develop physically motivated categories for astronomical objects. For example, it is worth considering whether the different types of terrestrial, volatile-rich, and gas giant planets should be clearly distinguished as highly distinct from each other. We might even consider the seemingly less obvious separation of stars above and below the Kraft break on the main sequence. Contrastingly, there is no clear distinction between gas giants and brown dwarfs, or between asteroids and comets. 

At the same time, places on the plots with no objects can also inform further investigations. There are no small objects far below the density of water, but there are several volatile worlds that are, prompting us to ask if there are ways to produce such small, underdense objects, or if it truly is impossible. The lack of overdense worlds beyond the mass boundary between volatile worlds and gas giants may provide limits on planetary formation conditions. 

However, these boundaries where objects suddenly stop being recorded may be due to a lack of information, rather than a true physical boundary--which shows another use of our work: to identify places where we lack the data to make robust conclusions. The regimes of dwarf planets and major moons are sparsely populated; the demographics for collapsing objects, mid-range brown dwarfs, and transitioning stars are limited, and we have no known exomoons. In time, we expect these regions to be filled as technology improves and surveys continue. However, we must be careful, as all small objects only have samples from our Solar System, and there is no reason to assume the distribution of objects in our Solar System is typical. 

Naturally, our plots can also be used in educational settings to show the relations between many different object classes in space all at once, rather than relying on students to draw the connections themselves as their studies move from topic to topic.
  
Future work could involve consideration of theoretical cohesive objects, such as Population III stars. Considering models, such as those for the tiny exoplanets we expect to exist but can't see, would also be worthwhile. There is also the consideration of objects that are not cohesive but are still distinct entities, such as protoplanetary discs, nebulae, globular clusters, and galaxies--these, too, have masses and densities. They could be assigned effective radii, though the concept of surface gravity would be somewhat nonsensical. Perhaps more connections between the objects in the universe are still waiting to be uncovered. 

Regardless, no matter what is done in the future or what parts of the cohesive object sequence are deemed the most interesting for study, we have at least shown how the properties of astronomical objects spanning more than thirty orders of magnitude in mass can be distilled into a single image with a bunch of colorful dots.

\vspace{\baselineskip}

The data used to create our graphs is available at the GitHub repository \url{https://github.com/GMBlackjack/CohesiveObjects/tree/main}. It includes the object's name, mass (in Earth masses), upper mass error, lower mass error, radius (in Earth radii), upper radius error, lower radius error, object classification, and sources used for the object in question. There are 2157 individual objects. 

We wish to draw attention to the sources that provided us with a particularly large number of data points, gathering the work of sometimes hundreds of other sources for us, easing the burden considerably. The NASA Exoplanet Archive \citep{NASAExoplanetArchive} provided almost all the exoplanets. The DEBCat catalog \citep{Southworth2014} had the largest list of stars we used, with significant additional contributions from both \cite{Torres2009} and \cite{Pineda2021}. \cite{Carry2012} was our primary source for asteroids.

Special mention goes to the NASA Planetary Physical Parameters and Planetary Satellite Physical Parameters web pages for gathering many sources used for the more well-known objects of the Solar System.

We also thank the anonymous reviewer for many helpful suggestions and insights.

\bibliography{Bibliography}{}
\bibliographystyle{aasjournal}

\end{document}